\documentclass{article}
\usepackage{template/spconf,amsmath,amssymb,hyperref,xcolor,graphicx}
\usepackage{caption}
\usepackage{subcaption}
\usepackage{cleveref}
\usepackage[absolute]{textpos}
\Crefname{figure}{Fig.}{Figs.}
\title{VSE: Variational state estimation of complex model-free process}

\name{Gustav Norén$^{\star \dagger}$ \quad
Anubhab Ghosh$^{\star}$ \quad
Fredrik Cumlin$^{\star}$  \quad
Saikat Chatterjee$^{\star}$
\thanks{The research is supported by funding from \href{https://www.digitalfutures.kth.se/}{Digital Futures Center}, \href{https://strategiska.se/}{Swedish Foundation for Strategic Research}, and \href{https://defence-industry-space.ec.europa.eu/system/files/2023-06/REACTII-Factsheet_EDF22.pdf}{European Defence Fund}.}}
\address{$^{\star}$School of Electrical Engineering and Computer Science, KTH Royal Institute of Technology, Sweden \\
      $^{\dagger}$Saab AB, Sweden \\
      \textit{gnoren@kth.se, anubhabg@kth.se, fcumlin@kth.se, sach@kth.se}}

\begin{document}
\ninept

\maketitle

\begin{textblock}{0.8}[0.5,0.5](0.5,0.955) 
\begin{center} \noindent © 2026 IEEE. Personal use of this material is permitted. Permission from IEEE must be obtained for all other uses, in any current or future media, including reprinting/republishing this material for advertising or promotional purposes, creating new collective works, for resale or redistribution to servers or lists, or reuse of any copyrighted component of this work in other works. 
\end{center}
\end{textblock}

\begin{abstract}
We design a variational state estimation (VSE) method that provides a closed-form Gaussian posterior of an underlying complex dynamical process from (noisy) nonlinear measurements. The complex process is model-free. That is, we do not have a suitable physics-based model characterizing the temporal evolution of the process state. The closed-form Gaussian posterior is provided by a recurrent neural network (RNN). The use of RNN is computationally simple in the inference phase. For learning the RNN, an additional RNN is used in the learning phase. Both RNNs help each other learn better based on variational inference principles. The VSE is demonstrated for a tracking application - state estimation of a stochastic Lorenz system (a benchmark process) using a 2-D camera measurement model. The VSE is shown to be competitive against a particle filter that knows the Lorenz system model and a recently proposed data-driven state estimation method that does not know the Lorenz system model.
\end{abstract}


\section{Introduction}
\label{sec:intro}

Bayesian state estimation of a complex dynamical process from noisy measurements is a challenging task. In this article, we deal with a complex process that is model-free. We do not have a suitable physics-based model to characterize the temporal dynamics of the process. Therefore, standard model-based methods like extended Kalman filter (EKF), unscented Kalman filter (UKF) \cite{ukf1997}, particle filters (PFs) have limitations for state estimation as they require a suitable model of the dynamical process. Instead, we require data-driven machine learning approaches. It is an additional challenge to design data-driven state estimation methods based on unsupervised learning using measurement-only training data.

To address the challenges, a \emph{data-driven nonlinear state estimation} (DANSE) method was proposed \cite{danse}. DANSE uses a linear measurement model and a recurrent neural network (RNN). The RNN provides a suitable Gaussian prior for the underlying state of the model-free process. Using the prior and the linear measurement model, DANSE provides a closed-form Gaussian posterior of the state. The learning of DANSE is unsupervised. The state estimation performance of DANSE was demonstrated for several stochastic nonlinear processes including a benchmark Lorenz system \cite{danse}. 

DANSE was extended for nonlinear measurements, referred to as particle-based DANSE (pDANSE) \cite{pdanse}. 
It uses a sampling-based technique to account for nonlinear measurements and provides second order statistics of the state posterior, but not a closed-form posterior. Further, it has a high computational cost in the inference phase due to the use of sampling to account for nonlinear measurements.  

In this article, we propose \emph{variational state estimation} (VSE)\footnote{Code is available at https://github.com/norengustav/vse.} method that provides a closed-form Gaussian posterior using nonlinear measurements, without sampling and incurring a high computational burden. The learning of VSE is unsupervised. In the learning phase, the VSE uses two RNNs: \(\text{RNN}_{\text{post}}\) and \(\text{RNN}_{\text{prior}}\). The two RNNs help each other learn better. Later, in the inference phase after the learning phase is over, only \(\text{RNN}_{\text{post}}\) is required. In the learning phase, \(\text{RNN}_{\text{prior}}\) provides a suitable Gaussian prior of the underlying state. Using the Gaussian prior and accounting nonlinear measurements, \(\text{RNN}_{\text{post}}\) learns to provide a closed-form Gaussian posterior based on variational inference. 
In the inference phase, \(\text{RNN}_{\text{prior}}\) is discarded and \(\text{RNN}_{\text{post}}\) is used directly to provide a Gaussian posterior as an approximation of the true posterior. The direct use of \(\text{RNN}_{\text{post}}\) to provide a Gaussian posterior avoids any sampling in the inference phase, saving computation massively. 

The VSE method is demonstrated for the Bayesian state estimation of the benchmark Lorenz system \cite{lorenzattractor} using a 2-D camera measurement model. It is compared with an unsupervised learning-based pDANSE and a PF that knows the Lorenz system model. Note that, given the camera measurement model and the Lorenz system model, the PF is an asymptotically optimal state estimation method that provides the best achievable performance as a performance bound. 

\noindent \textbf{Relevant literature:}
Data-driven methods exist in the literature for state estimation. A class of methods comprises dynamic variational autoencoders (DVAEs). They deal with model-free processes and provide posterior distributions using variational inference in unsupervised learning settings \cite{girin2021dynamical}. DVAEs were originally proposed for time-series modeling. Some of them can be adapted for state estimation, such as the deep Markov model (DMM) \cite{krishnan2017structured}. A limitation is that they are computationally intensive due to ancestral sampling, making them prohibitive for use in the case of longer sequence lengths and tracking applications. Another class of methods includes differentiable particle filters (DPFs). DPFs are learnable versions of traditional PFs \cite{jonschkowski2018differentiable, corenflos2021differentiable, chen2023dpfsurvey}. They use neural network-based Markovian models of the underlying dynamical process. The parameters of neural networks are learned in supervised and unsupervised settings \cite{jonschkowski2018differentiable, corenflos2021differentiable, van2025deep}. Lastly, we have hybrid approaches such as KalmanNet and its variations that explicitly use full/partial model of the dynamical process and specific operations (such as computing Kalman gain) are performed using deep neural networks \cite{revach2022kalmannet, revach2022unsupervised}. 

\section{Problem formulation and background}
\label{sec:problem-formulation}

Let $\mathbf{x}_{t} \in \mathbb{R}^{m}$ be the state vector of a dynamical process, where $t \in \mathbb{Z}_{\geq 0}$ denotes a discrete-time index. A $T$-length sequence generated from the dynamical process is compactly denoted as $\mathbf{x}_{1:T} \triangleq \mathbf{x}_{1}, \mathbf{x}_{2}, \ldots, \mathbf{x}_{t}, \ldots, \mathbf{x}_{T}$. The state is observed through a measurement system model, where $\mathbf{y}_{t} \in \mathbb{R}^{n}$ is a noisy measurement vector, as follows
\begin{equation}
\label{eq:measurementsys}
\mathbf{y}_{t} = \mathbf{h}(\mathbf{x}_{t}) + \mathbf{w}_{t}, \,\, \mathrm{for} \,\, t= 1,2, \ldots, T.
\end{equation}
In \cref{eq:measurementsys}, \(\mathbf{h}: \mathbb{R}^{m} \to \mathbb{R}^{n}\) represents a measurement system, for example, a camera measurement system; $\mathbf{w}_{t} \in \mathbb{R}^{n}$ denotes the measurement noise with known Gaussian distribution $p\left(\mathbf{w}_{t}\right) = \mathcal{N}\left(\mathbf{w}_{t}; \boldsymbol{0}, \mathbf{C}_{w} \right)$. Throughout this article, we assume that the measurement system model \cref{eq:measurementsys} is known; we know the function $\mathbf{h}$ and the Gaussian noise covariance $\mathbf{C}_{w}$. The measurement system model can be alternatively presented as $p(\mathbf{y}_t|\mathbf{x}_t) = \mathcal{N}(\mathbf{y}_t;\mathbf{h}(\mathbf{x}_{t}), \mathbf{C}_{w})$.

The Bayesian state estimation (BSE) task is to find an estimate of $\mathbf{x}_{t}$ using the sequence of past and present measurements $\mathbf{y}_{1:t} \triangleq \mathbf{y}_{1}, \mathbf{y}_{2}, \hdots \mathbf{y}_{t}$. That means estimating parameters of the posterior $p(\mathbf{x}_{t}|\mathbf{y}_{1:t}), \forall t$, or estimating statistical moments of $p(\mathbf{x}_{t}|\mathbf{y}_{1:t})$. This is a filtering problem. Note that we maintain causality, and the BSE can be used for many applications, for example, tracking of the dynamical process from noisy measurements.

For BSE, model-driven methods (such as EKF, UKF, PF, etc.) typically use a state-transition model (STM) of the complex dynamical process. An STM characterizes the temporal dynamics of the process. A widely used STM is Markovian with additive process noise, as follows 
\begin{equation}\label{eq:MarkovianNLDynamicalModel}
\mathbf{x}_{t} = \mathbf{f}(\mathbf{x}_{t-1}) + \mathbf{e}_{t}.
\end{equation}
Here, $\mathbf{f}: \mathbb{R}^{m} \to \mathbb{R}^{m}$ represents a relationship between the current state $\mathbf{x}_{t}$ and the previous state $\mathbf{x}_{t-1}$, and $\mathbf{e}_{t} \in \mathbb{R}^{m}$ represents the process noise to model stochasticity. In the relevant literature, the measurement system \cref{eq:measurementsys} and the STM \cref{eq:MarkovianNLDynamicalModel}  are together called the state-space model (SSM). The STM is used as a-priori knowledge for the model-driven methods.  In this article, the dynamical process is assumed to be model-free. We do not have any knowledge of the STM of the process; we do not know $\mathbf{f}$ and $\mathbf{e}_{t}$. Even we do not know whether a Markovian STM is good to model complex dynamics. Without an STM, the model-driven methods have limitations and we address the BSE in a data-driven manner that uses machine learning techniques. 

The use of machine learning requires training data. We have access to a training dataset consisting of $N$ independent measurement sequences. The training dataset is
    \(\mathcal{D} = \left\{ \mathbf{y}_{1:T}^{(i)}  \right\}_{i=1}^N,\)
where $\mathbf{y}_{1:T}^{(i)} \triangleq \mathbf{y}_{1}^{(i)}, \mathbf{y}_{2}^{(i)}, \hdots, \mathbf{y}_{T}^{(i)}$ is the $i$'th measurement sequence corresponding to the $i$'th state sequence $\mathbf{x}_{1:T}^{(i)}$. The $N$ measurement sequences can vary in length, for example, the $i$'th sequence can have $T^{(i)}$ length. We use the measurement sequences with the same $T$-length for notational clarity without loss of generality.

In a previous method DANSE \cite{danse} a linear measurement model is used, where the measurement system model \cref{eq:measurementsys} takes the form $\mathbf{y}_{t} = \mathbf{h}(\mathbf{x}_{t}) + \mathbf{w}_{t} = \mathbf{H}\mathbf{x}_{t} + \mathbf{w}_{t}$. While pDANSE \cite{pdanse} was developed for nonlinear measurement models. DANSE and pDANSE both use an RNN to provide a Gaussian prior $p(\mathbf{x}_t|\mathbf{y}_{1:t-1})$. Then, using the prior and the relevant measurement system model, they endeavor to compute the posterior $p(\mathbf{x}_t|\mathbf{y}_{1:t})$. Due to using a linear measurement model, DANSE provides a closed-form Gaussian posterior. On the other hand, pDANSE can not provide a closed-form posterior due to the use of a nonlinear measurement model, but it can provide second-order statistics based on a Monte-Carlo sampling technique.

Let the RNN corresponding to the prior for either DANSE or pDANSE have a set of parameters $\pmb{\theta}$. This RNN can be learned using the training dataset $\mathcal{D}$ in a maximum-likelihood approach by \(\max_{\pmb{\theta}} \log p(\mathcal{D})\) and in more detail:
\begin{equation}
    \max_{\pmb{\theta}} \displaystyle\sum_{i=1}^N \log p(\mathbf{y}_{1:T}^{(i)}) = \max_{\pmb{\theta}} \displaystyle\sum_{i=1}^N \sum_{t=1}^T \log p(\mathbf{y}_{t}^{(i)}|\mathbf{y}_{1:t-1}^{(i)}).
    \label{eq:MaximumLikelihoodOptimizationProblem}
\end{equation}
In the above, the chain rule of probability is used. For DANSE, we can compute $\log p(\mathbf{y}_{t}^{(i)}|\mathbf{y}_{1:t-1}^{(i)})$ in closed-form, and address the optimization problem \eqref{eq:MaximumLikelihoodOptimizationProblem} directly. On the other hand, for pDANSE, a sampling-based approach is used to compute a lower bound of $\log p(\mathbf{y}_{t}^{(i)}|\mathbf{y}_{1:t-1}^{(i)})$. Then the lower bound of the optimization cost $\sum_{i=1}^N \sum_{t=1}^T \log p(\mathbf{y}_{t}^{(i)}|\mathbf{y}_{1:t-1}^{(i)})$ is maximized. For both DANSE and pDANSE, the optimization requires a gradient search.

\begin{figure*}[ht]
\centering
\begin{subfigure}[b]{0.24\textwidth}
    \includegraphics[width=\textwidth]{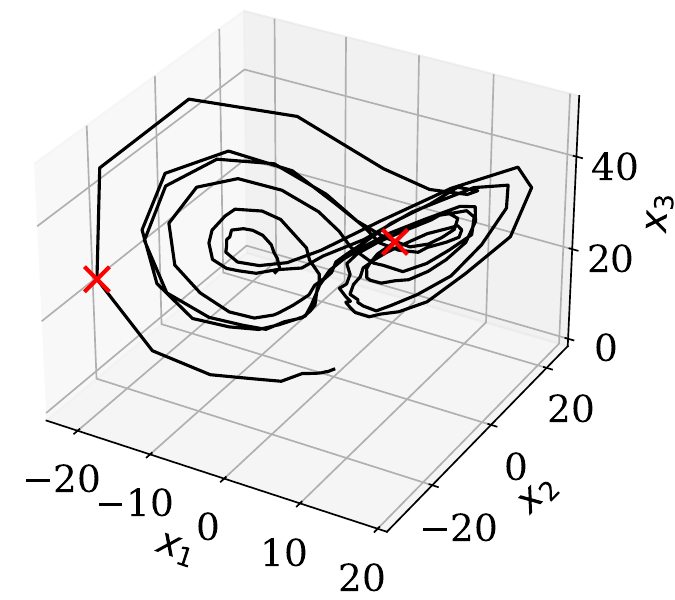}
    \caption{Trajectory of true state $\mathbf{x}_t$}
\end{subfigure}
\begin{subfigure}[b]{0.24\textwidth}
    \includegraphics[width=\textwidth]{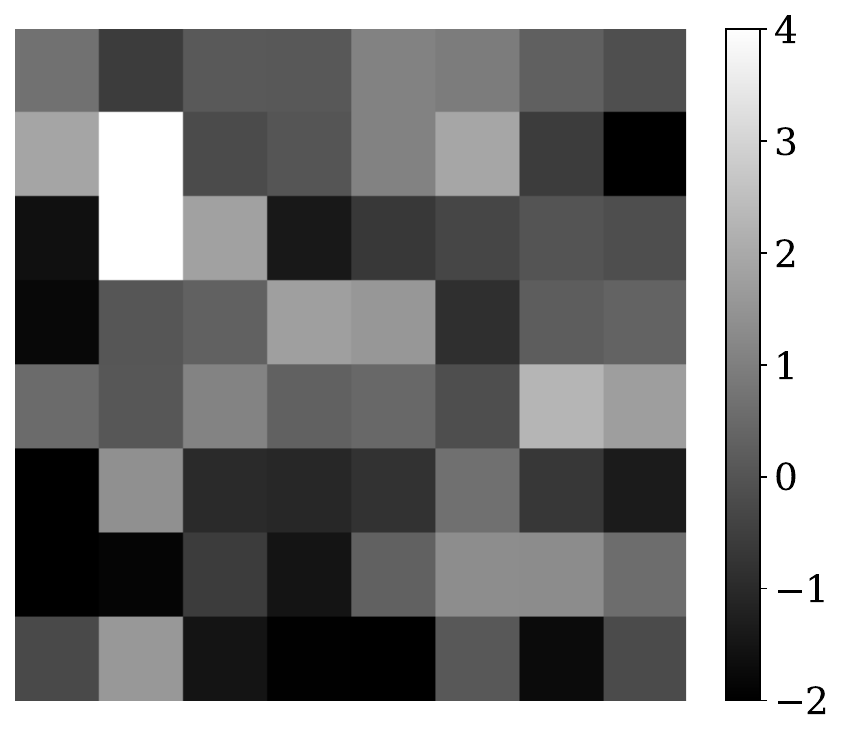}
    \caption{Image measurement $\mathbf{y}_t$ at \(t=10\)}
\end{subfigure}
\begin{subfigure}[b]{0.24\textwidth}
    \centering
    \includegraphics[width=\textwidth]{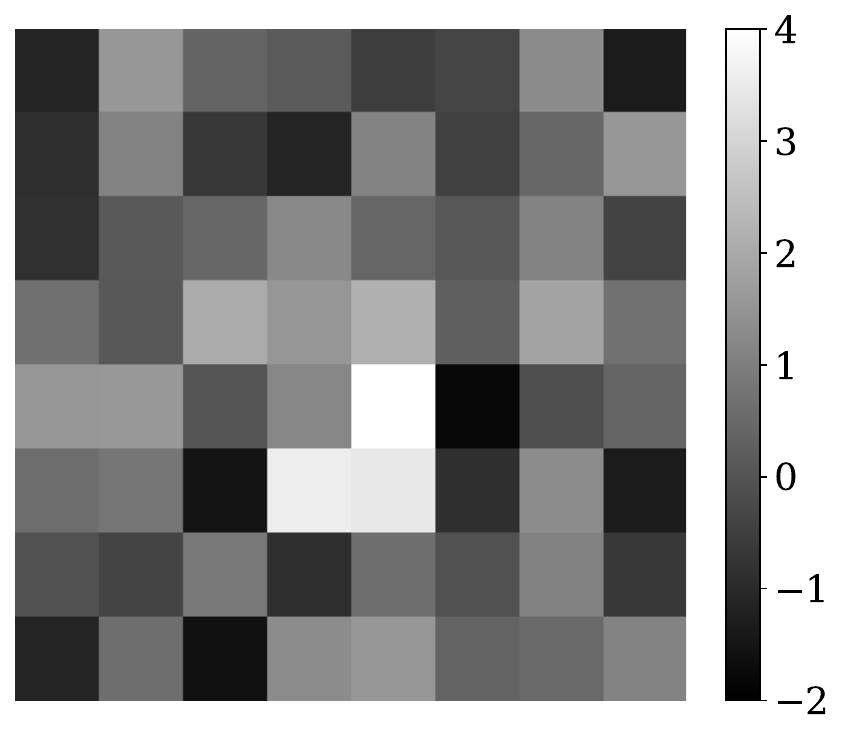}
    \caption{Image measurement $\mathbf{y}_t$ at \(t=50\)}
\end{subfigure}
\begin{subfigure}[b]{0.24\textwidth}
    \centering
    \includegraphics[width=\textwidth]{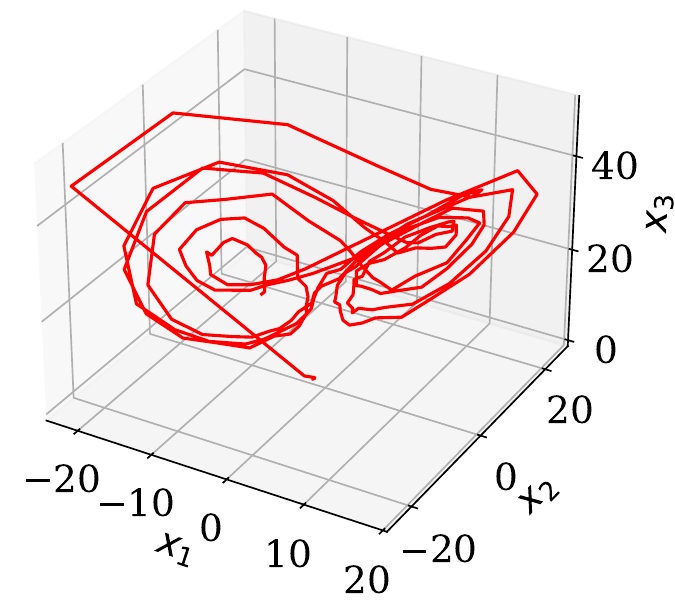}
    \caption{Estimated trajectory of $\hat{\mathbf{x}}_t$}
\end{subfigure}
\caption{Illustrating state estimation of VSE using camera model, at SMNR = 0 dB. (a) True state trajectory of a stochastic Lorenz system and red `$\times$' markers denote the 3-D coordinate positions at \(t=10\) and 50. (b) and (c) Images (measurements) captured using the camera model at \(t=10\) and 50. The images help to visualize the video as a sequence of images over time. (d) Estimated state trajectory of VSE.}
\label{fig:camera-example}
\end{figure*}

\section{VSE: Variational State Estimation}
\label{sec:VSE}

Often it is analytically intractable to compute the true posterior $p(\mathbf{x}_t|\mathbf{y}_{1:t})$ with a nonlinear measurement system model following \cref{eq:measurementsys}. We typically do not know $p(\mathbf{x}_t|\mathbf{y}_{1:t})$. 

In the proposed VSE method, instead of dealing with $p(\mathbf{x}_t|\mathbf{y}_{1:t})$, we endeavor to compute a Gaussian posterior in a variational sense. The Gaussian posterior is a proposal distribution, denoted by $q(\mathbf{x}_t|\mathbf{y}_{1:t}) = \mathcal{N}(\mathbf{x}_t; \mathbf{m}_{t|1:t}(\pmb{\psi}), \mathbf{L}_{t|1:t}(\pmb{\psi}))$. Here, $\mathbf{m}_{t|1:t}(\pmb{\psi})$ and $\mathbf{L}_{t|1:t}(\pmb{\psi}))$ are the mean vector and covariance matrix of the (variational) Gaussian posterior. We have a RNN that uses the measurement sequence $\mathbf{y}_{1:t}$ recursively and provides the Gaussian posterior parameters directly at each time point. The RNN is denoted by \(\text{RNN}_{\text{post}}\), with parameters $\pmb{\psi}$. That is why the parameters $\mathbf{m}_{t|1:t}(\pmb{\psi})$ and $\mathbf{L}_{t|1:t}(\pmb{\psi})$ are shown in terms of $\pmb{\psi}$. Let us, for now, assume that the learning phase of \(\text{RNN}_{\text{post}}\) is over. In the inference phase, \(\text{RNN}_{\text{post}}\) uses $\mathbf{y}_{1:t}$ directly and provides the variational posterior as follows:
\begin{eqnarray}
\label{eq:inference-phase}
\begin{array}{r}
    q(\mathbf{x}_t|\mathbf{y}_{1:t}) \triangleq q(\mathbf{x}_t|\mathbf{y}_{1:t};\pmb{\psi})) 
    = \mathcal{N}(\mathbf{x}_t; \mathbf{m}_{t|1:t}(\pmb{\psi}), \mathbf{L}_{t|1:t}(\pmb{\psi})),\\
    \text{such that } \text{RNN}_{\text{post}}(\mathbf{y}_{1:t}; \pmb{\psi}) \rightarrow \{\mathbf{m}_{t|1:t}(\pmb{\psi}), \mathbf{L}_{t|1:t}(\pmb{\psi})\}.\\
\end{array}
\end{eqnarray}
We have several remarks on the proposed VSE: (a) the variational posterior $q(\mathbf{x}_t|\mathbf{y}_{1:t})$ is a function of $\pmb{\psi}$ and that is why we express $q(\mathbf{x}_t|\mathbf{y}_{1:t}) \triangleq q(\mathbf{x}_t|\mathbf{y}_{1:t};\pmb{\psi}))$; (b) in the inference phase, \(\text{RNN}_{\text{post}}\) does not explicitly use the measurement model \eqref{eq:measurementsys} unlike DANSE and pDANSE; (c) the direct use of the RNN in the inference phase does not require sampling, saving a high computation requirement; (d) a point estimate of the state $\mathbf{x}_t$ can be $\hat{\mathbf{x}}_t=\mathbf{m}_{t|1:t}(\pmb{\psi})$. 

A natural question: How do we use the (nonlinear) measurement model \eqref{eq:measurementsys} and learn the \(\text{RNN}_{\text{post}}\) parameters $\pmb{\psi}$ that influence the inference phase? This is described in the next subsection.

\subsection{Unsupervised learning based on variational inference}

For learning \(\text{RNN}_{\text{post}}\), we use another RNN denoted by \(\text{RNN}_{\text{prior}}\). \(\text{RNN}_{\text{prior}}\) uses measurement sequence $\mathbf{y}_{1:t-1}$ to provide a Gaussian prior at the $t$'th time point. The use of \(\text{RNN}_{\text{prior}}\) for providing a Gaussian prior is similar to the use of an RNN in DANSE and/or pDANSE, although with a major difference. For VSE, \(\text{RNN}_{\text{prior}}\) is used in the learning phase only and not in the inference phase. On the other hand, for DANSE and pDANSE, the RNN that provides the Gaussian prior is used in both the learning and inference phases. 

Like DANSE and pDANSE notations, let \(\text{RNN}_{\text{prior}}\) have the set of parameters denoted by $\pmb{\theta}$. \(\text{RNN}_{\text{prior}}\) provides the prior $p(\mathbf{x}_t|\mathbf{y}_{1:t-1}) = \mathcal{N}(\mathbf{x}_t; \mathbf{m}_{t|1:t-1}(\pmb{\theta}), \mathbf{L}_{t|1:t-1}(\pmb{\theta}))$. Here, $\mathbf{m}_{t|1:t-1}(\pmb{\theta})$ and $\mathbf{L}_{t|1:t-1}(\pmb{\theta}))$ are the mean vector and covariance matrix of the Gaussian prior. \(\text{RNN}_{\text{prior}}\) uses the measurement sequence $\mathbf{y}_{1:t-1}$ recursively to provide the prior mean and covariance. The measurement model and the Gaussian prior together are as follows:
\begin{eqnarray}
\label{eq:learning-phase}
\begin{array}{l}
\begin{array}{l}
     p(\mathbf{y}_t|\mathbf{x}_t) = \mathcal{N}(\mathbf{y}_t;\mathbf{h}(\mathbf{x}_t), \mathbf{C}_w); 
\end{array}
      \\
     \begin{array}{ll}
     p(\mathbf{x}_t|\mathbf{y}_{1:t-1}) \!\!\!\! & \triangleq p(\mathbf{x}_t|\mathbf{y}_{1:t-1};\pmb{\theta}) \\
     \!\!\!\! & = \mathcal{N}(\mathbf{x}_t; \mathbf{m}_{t|1:t-1}(\pmb{\theta}), \mathbf{L}_{t|1:t-1}(\pmb{\theta})),
     \end{array} \\
    \begin{array}{r}
    \text{such that } \text{RNN}_{\text{prior}}(\mathbf{y}_{1:t-1}; \pmb{\theta}) \! \rightarrow \! \{\! \mathbf{m}_{t|1:t-1}(\pmb{\theta}), \! \mathbf{L}_{t|1:t-1}(\pmb{\theta}) \! \}.  
    \end{array}
\end{array}
\end{eqnarray}
Note that the prior $p(\mathbf{x}_t|\mathbf{y}_{1:t-1})$ is a function of $\pmb{\theta}$ and that is why we write $p(\mathbf{x}_t|\mathbf{y}_{1:t-1})  \triangleq p(\mathbf{x}_t|\mathbf{y}_{1:t-1};\pmb{\theta})$.

Our task is to learn the parameters $\pmb{\theta}$ and $\pmb{\psi}$ of the two RNNs in a maximum-likelihood optimization principle as follows:
\begin{equation}
\label{eq:MaximumLikelihoodOptimizationProblem_VSE}
     \max_{\pmb{\theta},\pmb{\psi}} \log p(\mathcal{D}) \implies \max_{\pmb{\theta},\pmb{\psi}} \displaystyle\sum_{i=1}^N \sum_{t=1}^T \log p(\mathbf{y}_{t}^{(i)}|\mathbf{y}_{1:t-1}^{(i)}).
\end{equation}
Let us drop the superscript $(i)$ from $\log p(\mathbf{y}_{t}^{(i)}|\mathbf{y}_{1:t-1}^{(i)})$ for notational clarity. 
The main idea is to maximize a lower bound  \(\mathcal{L}_t(\mathbf{y}_{1:t}; \pmb{\theta}, \pmb{\psi})\) of the logarithmic evidence $\log p(\mathbf{y}_t | \mathbf{y}_{1:t-1})$.
The derivation is as follows:
\begin{equation}
\begin{array}{l}
    \log p(\mathbf{y}_t | \mathbf{y}_{1:t-1}) \\
    = \int_{\mathbf{x}_t} q (\mathbf{x}_{t}|\mathbf{y}_{1:t}) \log p(\mathbf{y}_t | \mathbf{y}_{1:t-1}) \, d\mathbf{x}_t \\
    = \int_{\mathbf{x}_t} q (\mathbf{x}_{t}|\mathbf{y}_{1:t}) \log \frac{p(\mathbf{x}_t| \mathbf{y}_t , \mathbf{y}_{1:t-1}) p(\mathbf{y}_t | \mathbf{y}_{1:t-1})}{p(\mathbf{x}_t| \mathbf{y}_t , \mathbf{y}_{1:t-1})} \, d\mathbf{x}_t \\
    = \int_{\mathbf{x}_t} q (\mathbf{x}_{t}|\mathbf{y}_{1:t}) \log\frac{ p(\mathbf{x}_t,\mathbf{y}_t | \mathbf{y}_{1:t-1})}{p(\mathbf{x}_t| \mathbf{y}_{1:t})} \, d\mathbf{x}_t \\
    = \underbrace{\displaystyle\int_{\mathbf{x}_t} q (\mathbf{x}_{t}|\mathbf{y}_{1:t}) \log \frac{p(\mathbf{x}_t,\mathbf{y}_t | \mathbf{y}_{1:t-1})}{q(\mathbf{x}_t| \mathbf{y}_{1:t})} \, d\mathbf{x}_t }_{\mathcal{L}_t(\mathbf{y}_{1:t})} \\
    \hspace{0.5cm} + \underbrace{\displaystyle\int_{\mathbf{x}_t} q (\mathbf{x}_{t}|\mathbf{y}_{1:t}) \log\frac{q(\mathbf{x}_t| \mathbf{y}_{1:t})}{p(\mathbf{x}_t| \mathbf{y}_{1:t})} \, d\mathbf{x}_t}_{\mathrm{D}_{\mathrm{KL}}(q\|p)} \\
    \geq \mathcal{L}_t(\mathbf{y}_{1:t}) \triangleq \mathcal{L}_t(\mathbf{y}_{1:t};\pmb{\theta},\pmb{\psi}).
\end{array}
\end{equation}
Here, $\mathrm{D}_{\mathrm{KL}}(q\|p)$ is the Kullback–Leibler (KL) divergence between $q(\mathbf{x}_t| \mathbf{y}_{1:t})$ and $p(\mathbf{x}_t| \mathbf{y}_{1:t})$. As $\mathrm{D}_{\mathrm{KL}}(q\|p) \geq 0$,  $\mathcal{L}_t(\mathbf{y}_{1:t})$ is a lower bound of logarithmic evidence. Maximization of $\mathcal{L}_t(\mathbf{y}_{1:t})$ leads to decrease in $\mathrm{D}_{\mathrm{KL}}(q\|p)$, and $q(\mathbf{x}_t| \mathbf{y}_{1:t})$ approaches  $p(\mathbf{x}_t| \mathbf{y}_{1:t})$.

We express the lower bound using the relation $p(\mathbf{x}_t,\mathbf{y}_t | \mathbf{y}_{1:t-1}) = p(\mathbf{y}_{t} | \mathbf{x}_t) \, p(\mathbf{x}_{t} | \mathbf{y}_{1:t-1}; \pmb{\theta})$ as follows:
\begin{eqnarray}
\begin{array}{rl}
\mathcal{L}_t(\mathbf{y}_{1:t};\pmb{\theta,\psi}) \!\!\!\!\!\! &
= \!\! \int_{\mathbf{x}_t} \!\! q (\mathbf{x}_{t}|\mathbf{y}_{1:t}; \! \pmb{\psi}) \log \!\! \frac{p(\mathbf{y}_{t} | \mathbf{x}_t) \, p(\mathbf{x}_{t} | \mathbf{y}_{1:t-1}; \pmb{\theta}) }{q(\mathbf{x}_t| \mathbf{y}_{1:t}; \pmb{\psi})}  d\mathbf{x}_t \\
& = \!\! \int_{\mathbf{x}_t} q (\mathbf{x}_{t}|\mathbf{y}_{1:t}; \pmb{\psi}) \log {p(\mathbf{y}_{t} | \mathbf{x}_t)} \, d\mathbf{x}_t \\ 
& \hspace{0.5cm}- \int_{\mathbf{x}_t} q (\mathbf{x}_{t}|\mathbf{y}_{1:t}; \pmb{\psi}) \log \frac{q(\mathbf{x}_t| \mathbf{y}_{1:t}; \pmb{\psi})}{p(\mathbf{x}_{t} | \mathbf{y}_{1:t-1}; \pmb{\theta}) } \, d\mathbf{x}_t \\
& = \mathbb{E}_{q (\mathbf{x}_{t}|\mathbf{y}_{1:t}; \pmb{\psi})}\left[\log {p(\mathbf{y}_{t} | \mathbf{x}_t)}\right] \\
& \hspace{0.5cm}- \mathrm{D}_{\mathrm{KL}}\left({q(\mathbf{x}_t| \mathbf{y}_{1:t}; \pmb{\psi})} \Vert {p(\mathbf{x}_{t} | \mathbf{y}_{1:t-1}; \pmb{\theta}) } \right).
\end{array}
\end{eqnarray}
In the above, $\mathbb{E}$ denotes statistical expectation operation. The final optimization problem is a maximization of the evidence lower bound for the training dataset $\mathcal{D}$, as follows:
\begin{equation}
    \max_{\pmb{\theta}, \pmb{\psi}} \sum_{i=1}^N \sum_{t=1}^{T} \mathcal{L}_t(\mathbf{y}^{(i)}_{1:t};\pmb{\theta,\psi}).
\end{equation}

For optimization, $\mathbb{E}_{q (\mathbf{x}_{t}|\mathbf{y}_{1:t}; \pmb{\psi})}\left[\log_{e} {p(\mathbf{y}_{t} | \mathbf{x}_t)}\right]$ is approximated by drawing $L$ samples from $q (\mathbf{x}_{t}|\mathbf{y}_{1:t}; \pmb{\psi})$ and numerically averaging the logarithmic measurement model $\log {p(\mathbf{y}_{t} | \mathbf{x}_t)}$. The KL divergence $\mathrm{D}_{\mathrm{KL}}\left({q(\mathbf{x}_t| \mathbf{y}_{1:t}; \pmb{\psi})} \Vert {p(\mathbf{x}_{t} | \mathbf{y}_{1:t-1}; \pmb{\theta}) } \right)$ involves two Gaussian distributions and has a closed-form expression. The sampling-based average and KL divergence expression are as follows:
\begin{eqnarray*}
\begin{array}{l}
\mathbb{E}_{q (\mathbf{x}_{t}|\mathbf{y}_{1:t}; \pmb{\psi})}\left[\log {p(\mathbf{y}_{t} | \mathbf{x}_t)}\right] {\approx} \displaystyle\frac{1}{L} \sum_{l=1}^{L} \log {\mathcal{N}(\mathbf{y}_t ; \mathbf{h}(\mathbf{x}_t^{(l)}\left(\pmb{\psi}\right)), \mathbf{C}_w ) }, \\
\text{where }\mathbf{x}_t^{(l)}\left(\pmb{\psi}\right) \sim q(\mathbf{x}_t| \mathbf{y}_{1:t}; \pmb{\psi}) = \mathcal{N}(\mathbf{x}_t; \mathbf{m}(\pmb{\psi}),  \mathbf{L}(\pmb{\psi})); \\
\mathrm{D}_{\mathrm{KL}}\left({q(\mathbf{x}_t| \mathbf{y}_{1:t}; \pmb{\psi})} \Vert {p(\mathbf{x}_{t} | \mathbf{y}_{1:t-1}; \pmb{\theta}) } \right) {=} \frac{1}{2}\bigg[\!\log \frac{\det\mathbf{L} \left(\pmb{\theta}\right)}{\det\mathbf{L} \left(\pmb{\psi}\right)} - m \\
+ \left[ \mathbf{m}\!\left(\pmb{\psi}\right) \! - \!{\mathbf{m}\!\left(\pmb{\theta}\right)} \right]^{\!\top} \! \mathbf{L}^{-1}({\pmb{\theta}}) \! \left[ \mathbf{m}\!\left(\pmb{\psi}\right) \! - \!{\mathbf{m}\!\left(\pmb{\theta}\right)} \right] 
+ \text{tr}\left({\mathbf{L}^{-1}({\pmb{\theta}})}\mathbf{L}\left(\pmb{\psi}\right)\right)\!\bigg].
\end{array}
\end{eqnarray*}
In the above, we compactly write \(\mathbf{m}_{t|1:t}(\pmb{\psi}) \triangleq \mathbf{m}(\pmb{\psi})\), \(\mathbf{m}_{t|1:t-1}(\pmb{\theta}) \triangleq \mathbf{m}(\pmb{\theta})\), \(\mathbf{L}_{t|1:t}(\pmb{\psi}) \triangleq \mathbf{L}(\pmb{\psi})\) and \(\mathbf{L}_{t|1:t-1}(\pmb{\theta}) \triangleq \mathbf{L}(\pmb{\theta})\) for brevity. We use a reparameterization trick to draw $L$ samples as $\mathbf{x}_t^{(l)}\left(\pmb{\psi}\right)
= \mathbf{m}_{t|1:t}(\pmb{\psi}) + \mathbf{L}^{\frac{1}{2}}_{t|1:t}(\pmb{\psi}) \pmb{\epsilon}^{(l)}$, where $\pmb{\epsilon}^{(l)} \sim \mathcal{N}\left(\pmb{\epsilon}; \boldsymbol{0}, \mathbf{I}_{m}\right)$ and $\mathbf{L}^{\frac{1}{2}}_{t|1:t}(\pmb{\psi})$ is the Cholesky factor of $\mathbf{L}_{t|1:t}(\pmb{\psi})$. The reparameterization trick efficiently performs a gradient search \cite{reparameterization}.

In the learning phase, \(\text{RNN}_{\text{prior}}\) and \(\text{RNN}_{\text{post}}\) help each other learn better. Once the learning phase is over, \(\text{RNN}_{\text{prior}}\) is discarded and only \(\text{RNN}_{\text{post}}\) is used in the inference phase.

\section{Experiments and Results}
\label{sec:experiments_and_results}
The proposed VSE is demonstrated to track a 3-D stochastic Lorenz system using a 2-D camera measurement model. The state estimation of a 3-D stochastic Lorenz system is a benchmark problem in the relevant literature. The VSE is compared with pDANSE and PF. The pDANSE uses unsupervised learning and does not know the stochastic Lorenz system, just like VSE. On the other hand, PF knows the stochastic Lorenz system and is an asymptotically optimal estimator providing the best achievable performance as a lower bound.

The dynamics of the stochastic Lorenz system with state $\mathbf{x}_t \triangleq [x_{t,1} \,\, x_{t,2} \,\, x_{t,3}]^{\top} \in \mathbb{R}^3$ is as follows:
\begin{equation}
\begin{aligned}
    \mathbf{x}_{t+1} & =  \mathbf{f}(\mathbf{x}_{t-1}) + \mathbf{e}_{t} = \mathbf{F}_t(\mathbf{x}_t)\mathbf{x}_t + \mathbf{e}_t \in \mathbb{R}^3, \\
    \mathbf{F}_t(\mathbf{x}_t) & = \exp \left(
    \begin{bmatrix}
        -10 & 10 & 0 \\
        28 & -1 & -x_{t,1} \\
        0 & x_{t,1} & -8/3
    \end{bmatrix}
    \Delta
    \right),
\end{aligned}
\label{eq:lorenz-ssm}
\end{equation}
where \(\mathbf{e}_t \sim \mathcal{N}(\mathbf{e}_t;\; \mathbf{0},\mathbf{C}_e)\) is the process noise with \(\mathbf{C}_e = \sigma^2_e\mathbf{I}_3\), step-size \(\Delta = 0.02 \text{ seconds}\), and \(\sigma^2_e\) chosen to correspond to \(-10\) dB. For simulations we use a 5'th order Taylor approximation for \(\mathbf{F}_t(\mathbf{x_t})\) in \cref{eq:lorenz-ssm}. This setup is similar to works in \cite{danse, revach2022kalmannet, garcia2019combining}. Note that while a standard Lorenz system (aka Lorenz attractor) is a chaotic dynamical system and deterministic in nature \cite{lorenzattractor}, we have the process noise $\mathbf{e}_t$ to bring stochasticity.

The 2-D camera measurement model is motivated by its use in \cite[eq. (26)]{camera-model}. The camera model uses a Gaussian point spread function over a predetermined grid representing pixels of a simulated camera. Let the camera be of low resolution; it has $8 \! \times \! 8 \! = \! 64$ pixels. The camera is placed in the 2-D plane with axes $x_{1}$ and $x_{2}$. It has equally spaced grid points in a rectangular plane that holds the pixels in said grid. The grid points \(\{ \mathbf{g}_i \}_{i=1}^{64}\) are in the range $[-30,30]$ along $x_{1}$ and $[-40,40]$ along $x_{2}$. Then the camera perceives $x_{3}$-axis as depth. The state vector $\mathbf{x}_t$ is projected in the camera plane. We can express the camera measurement function in its vectorized form as $\mathbf{h}(\mathbf{x}_t) \triangleq [h_{t,1}, \, h_{t,2}, \ldots, h_{t,64}]^{\top} \in \mathbb{R}^{64}$, where $h_{t,i}$ denotes the intensity of the $i$'th pixel which is modeled as follows:
    \begin{equation}
    h_{t,i} = 10 \exp \left(- \frac{1}{2x_{t,3}} \left\| \mathbf{g}_i -
    \begin{bmatrix}
        x_{t,1} \\
        x_{t,2}
    \end{bmatrix}
    \right\|^2_2 \right).
\end{equation}
In the above, the exponent function represents the Gaussian spread function. The simulated 2-D camera model is an \(\ell_2\)-distance based system and susceptible to phase flip of the state $\mathbf{x}_t$. It does not preserve the phase information for the following reason: with respect to the camera position, a 3-D coordinate position of \(\mathbf{x}_t\) and its mirror image coordinate position have the same distances from the camera.

It is possible to simulate a high-resolution camera by having more pixels. We choose a modest pixel size of \(8\times8\) representing a cheap (and noisy) camera, making the estimation task challenging. 
With the additive Gaussian noise to the pixels, the camera  measurement system follows \cref{eq:measurementsys}, where \(\mathbf{w}_t \sim \mathcal{N}(\mathbf{w}_t;\; \mathbf{0},\mathbf{C}_w)\) with $\mathbf{C}_w=\sigma^2_w \mathbf{I}_{64}$ and $\sigma^2_w$ denotes the variance of measurement noise. By choosing a value of \( \sigma^2_w \), for $N$ measurement sequences, we have the signal-to-measurement-noise-ratio (SMNR) in dB as follows:
\begin{equation*}
    \mathrm{SMNR} {=}  \frac{1}{N}\!\sum_{i=1}^{N} 10\log_{10}\sum_{t=1}^{T}\frac{\mathbb{E}\!\left[\!\left|\!\left|\mathbf{h}(\mathbf{x}_t^{(i)})\!-\mathbb{E}\left[{\mathbf{h}(\mathbf{x}_t^{(i)})}\right]\!\right|\!\right|^2_2\right]}{\text{tr}(\mathbf{C}_w)}.
\end{equation*}
A high $\sigma^2_w$ leads to a low SMNR. 
A (noisy) measurement sequence is a video observing a state trajectory of the stochastic Lorenz system. \Cref{fig:camera-example} shows an illustration of a state trajectory, two images captured by the camera at two time points of the corresponding video sequence, and the sequence estimated by VSE.

VSE and pDANSE are trained using the same training dataset $\mathcal{D}$ consisting of \(N_{train}=1000\) sequences of length \(T_{train} = 200\). The test dataset consists of \(N_{test}=100\) sequences of length \(T_{test} = 1000\). Training and test datasets are generated separately for varying SMNRs. To evaluate experimental results, the performance measure is normalized-mean-square-error (NMSE) adjusted to invariance in phase flips. 
Similar considerations and details on phase flips can be found in the literature \cite{candes2013phaselift}. The NMSE in dB is:
\begin{equation*}
    \mathrm{NMSE} = \frac{1}{N_{\text{test}}}\sum_{i=1}^{N_{\text{test}}}10\log_{10} \frac{\sum_{t=1}^{T^{(i)}} \|\mathrm{abs}_e(\mathbf{x}_t^{(i)})-\mathrm{abs}_e(\hat{\mathbf{x}}_t^{(i)}) \|^2_2}{\sum_{t=1}^{T^{(i)}} \|\mathbf{x}_t^{(i)} \|_2^2},
\end{equation*}
where \(\mathrm{abs}_e(\cdot)\) performs element-wise absolute value operation addressing phase invariance of the state. A low NMSE is desirable.


\vspace{0.3cm}
\noindent{\textbf{Training details:}} For pDANSE and VSE, we used gated recurrent units (GRUs) as RNNs \cite{choPropertiesNeuralMachine2014}. In our implementation, a GRU has two hidden layers with 80 hidden units. The GRU output is mapped via a fully-connected layer of 128 hidden units to mean and covariance parameters. The GRU combined with the fully-connected layer network architecture was chosen by an experimental validation. VSE and pDANSE were trained using Adam optimizer \cite{kingma2014adam}. We used a mini-batch gradient descent with batch size 128, 500-to-1000 epochs, and an adaptive learning rate starting at \(10^{-3}\). The VSE used $L=10$ samples for the reparameterization trick. The pDANSE used 100 particles (samples) in its simulations. All simulations were performed using Python and PyTorch \cite{paszke2019pytorch} and run on the same computer having a GPU.

\vspace{0.3cm}
\noindent{\textbf{Results:}}
\begin{figure}[t]
    \centering
    \includegraphics[width=1\linewidth]{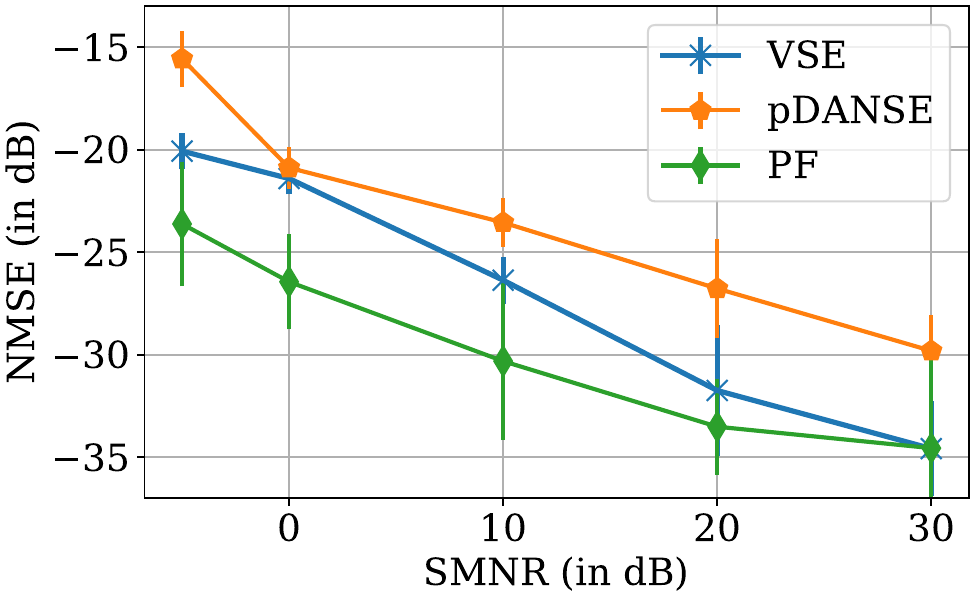}
    \caption{NMSE versus SMNR performances of VSE, pDANSE and PF. The PF provides the best achievable performance.}
    \label{fig:nmse_plot_highdim}
\end{figure}
The NMSE versus SMNR performances for VSE, pDANSE and PF are shown in \Cref{fig:nmse_plot_highdim}. VSE shows better performance than pDANSE. PF provides the best achievable performance, and there exists a gap between PF and VSE. We now provide a comparison between computational requirements. In the inference phase, the total simulation times for pDANSE and VSE are 41.5 and 0.246 seconds, respectively. VSE is significantly faster than pDANSE and provides better state estimation performance.

\section{Conclusion}
\label{sec:conclusion}
The proposed VSE method efficiently addresses Bayesian state estimation of a model-free process from nonlinear measurements in a unsupervised learning setting. Using variational inference principles, two recurrent neural networks learn from measurement-only dataset. In the inference phase, a relevant RNN uses nonlinear measurement sequence directly maintaining causality and provides a closed-form Gaussian posterior without incurring a high computational complexity. In the inference phase, the method is sampling-free. Future works include experimenting with various nonlinear measurement models and processes.

\clearpage
\bibliographystyle{template/IEEEbib}
\bibliography{refs}

\end{document}